\documentclass[aps,prb,twocolumn,showpacs]{revtex4-1}

\usepackage{amsfonts,amssymb,amsmath}
\usepackage{mathtools}
\usepackage[]{graphics,graphicx,epsfig}
\usepackage{amsthm,multirow}
\usepackage{color}
\begin{document}

\title{Coherent multi-mode conversion from microwave to optical wave via a magnon-cavity hybrid system}

\author{Yong Sup \surname{Ihn}}

\author{Su-Yong Lee}

\author{Dongkyu Kim}

\author{Sin Hyuk Yim}

\author{Zaeill \surname{Kim}}
\email{corresponding: z.kim@add.re.kr}

\affiliation{Quantum Physics Technology Directorate, Agency for Defense Development, Daejeon 34186, Korea}

\date{\today}

\begin{abstract}
Coherent conversion from microwave to optical wave opens new research avenues towards long distant quantum network covering quantum communication, computing, and sensing out of the laboratory.
Especially multi-mode enabled system is essential for practical applications.
Here we experimentally demonstrate coherent multi-mode conversion from the microwave to optical wave via collective spin excitation in a single crystal
yttrium iron garnet (YIG, Y$_{3}$Fe$_{5}$O$_{12}$) which is strongly coupled to a microwave cavity mode in a three-dimensional rectangular cavity.
Expanding collective spin excitation mode of our magnon-cavity hybrid system from Kittel to multi magnetostatic modes, we verify that the size of YIG sphere predominantly plays a crucial role for the microwave-to-optical multi-mode conversion efficiency at resonant conditions.
We also find that the coupling strength between multi magnetostatic modes and a cavity mode is manipulated by the position of a YIG inside the cavity.
It is expected to be valuable for designing a magnon-hybrid system that can be used for coherent conversion between microwave and optical photons.

\end{abstract}

\maketitle

\section{Introduction}
Strong coupling induced by resonant light-matter interaction can give rise to coherent information transfer between distinct physical systems in quantum and classical information processing \cite{Nature08Kimble,PRL97Cirac,PRL90Spreeuw}. The coherent transfer of quantum state is a key role in realizing large-scale quantum optical networks and long distance quantum sensing and imaging \cite{PRL13Julsgaard,CRP16Greze,QST17Sasaki,PS09Wallquist,APL19Chang,NCOM19Pogorzalek} since it allows quantum information to be exchanged between different systems that operate at different energy scales. 
A platform for transferring multi-mode states will be attractive to the practical application of quantum-enhance metrology and communication \cite{Nature13Vincenzo,Nature15Becerra}.
After the first demonstration of optical frequency conversion \cite{PRL92Huang}, the photon frequency conversion has been implemented with crystals in optical domain, and with superconducting circuits in the microwave domain \cite{PRL13Abdo,PRL16Lecocq}.
Since it has great potential in realizing large-scale quantum optical networks with superconducting qubits, recently,
the microwave to optical field conversion has been intensively attracted and experimentally demonstrated by using intermediate systems, such as optomechanical systems \cite{PRL10Stannigel,NP14Andrew,NP18Higginbotham,NP14Bagci,PRL12Barzanjeh,PRL12Wang}, electro-optical systems \cite{PRA17Soltani}, atomic ensembles \cite{PRL09Imamoglu,PRL09Verdu,PRL14O'Brien,PRL14Williamson}, and magnons \cite{PRB16Hisatomi}.
So far, the maximum microwave-to-optical (MO) conversion efficiency has been demonstrated in a nanomechanical resonator system employing a nano-membrane that is combined with an optical cavity while it is coupled to a superconducting microwave resonator. 
Its conversion efficiency reached 47 $\%$ at low temperature \cite{NP14Andrew,NP18Higginbotham}. A ferromagnetic material, an yttrium iron garnet (YIG), in a microwave cavity offers strong interaction between magnon and microwave cavity modes at both low and room temperatures since YIG has a Curie temperature of 560 K and a net spin density of $2.1 \times 10^{22}$$\mu_{B}$cm$^{-3}$ ($\mu_{B}$; Bohr magneton) that is few orders of magnitude higher than a net spin density of  10$^{16}-$10$^{18}$$\mu_{B}$cm$^{-3}$ in paramagnetic materials.
High Verdet constant with sharp linewidth of electron spin resonance in microwave domain also makes YIG noticeable in Faraday effect \cite{OFST15Grattan,15Belov}. Recently, YIG based materials have been studied on a novel concept for ultrafast magneto-optic polarization modulation with frequencies up to THz orders \cite{16Walowski,20Maccaferri}.
Longer spin excitation time than paramagnetic spin system is another advantage of YIG and its hybrid system \cite{PRL14Tabuchi,PRL14Zhang}.
In this hybrid system, the MO conversion is achieved through the Faraday effect and Purcell effect. 
The magnetization oscillation induced and amplified by the Purcell effect of a microwave cavity mode creates the sidebands to the incidental optical wave, resulting in coherent conversion between microwave and optical wave.
So far, the MO conversion via YIG-cavity system has been focused only for the Kittel mode.

In this paper, we report coherent multi-mode conversion from the microwave to optical wave fields, which is based on a hybrid system consisting of a sphere of YIG and a three-dimensional rectangular microwave cavity. We experimentally demonstrate and verify that the size of YIG is a dominant factor of the coherent multi-mode conversion efficiency. The conversion efficiency is theoretically derived by using the interaction Hamiltonian with the ferromagnetic-resonance (FRM or Kittel mode, KM) and a higher magnetostatic mode (MSM) and experimentally characterized by normal-mode splitting, coupling strengths of the ferromagnetic-resonance, and magnetostatic modes. For the near-uniform microwave cavity field, all the spins in the YIG sphere precess in phase that is called Kittel mode, and therefore the whole YIG sphere can be treated as a giant spin \cite{PR47Kittel,PR48Kittel}. As the YIG sphere size increases, the microwave cavity field can no longer be treated as a uniform field, leading to the higher order magnon modes which are observed \cite{PR57Walker,PR58Walker,JAP59Fletcher}. 
By positioning a YIG off the uniform microwave cavity field region, we can manipulate the coupling strength between multi magnetostatic modes and a cavity.
As a result, it is shown that KM and MSM manifested in each YIG sample are successfully transferred through the conversion process from microwave field to optical wave field.
\begin{figure}
\centerline{\scalebox{0.8}{\includegraphics[angle=0]{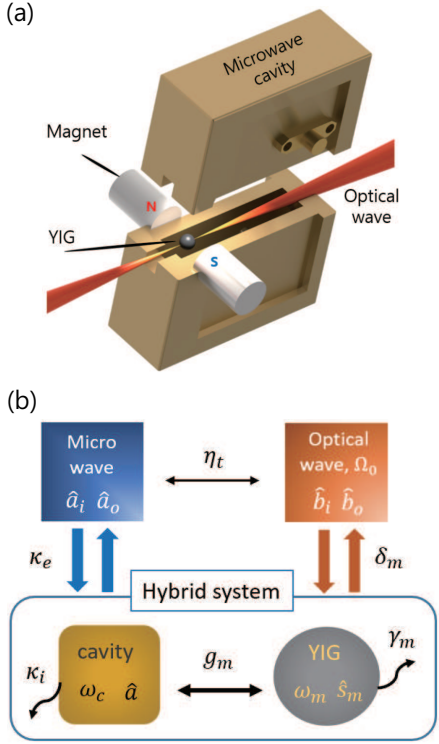}}}
\caption{(a) Schematic of hybrid system for coherent coversion from microwave to light, where an YIG sphere is inserted into a rectangular microwave cavity. (b) Magnon-cavity hybrid model. In the hybrid system, magnon mode $\hat{s}_{m}$ and a microwave cavity mode $\hat{a}$ are strongly coupled with a coupling strength $g_{m}$. Here, $\kappa_{i}$ and $\gamma_{m}$ present internal cavity loss and intrinsic loss for magnon modes, respectively. A microwave field mode $\hat{a}_{i}$ is coupled to a microwave cavity mode at a rate $\kappa_{e}$, while a traveling optical wave field mode $\hat{b}_{i}$ is coupled to the magnon mode at a rate $\delta_{m}$. Through this process, microwave field is converted to the traveling optical wave field with a conversion efficiency $\eta_{t}$.}
\label{fig1}
\end{figure}

\section{Theoretical Conversion Model}
Fig. \ref{fig1}(a) shows the main part of our hybrid system, a YIG sphere and a 3D rectangular microwave cavity. 
Due to the magnetic and optical properties, a highly polished YIG sphere can serve as an excellent magnon resonator at microwave frequencies.
A 3D rectangular microwave cavity intrinsically maintains a low damping rate compatible with one of magnon mode at room temperature and enhances the coupling rate between a magnon mode and a cavity mode through the Purcell effect.
Then, the linearly polarized light travels through a YIG perpendicular to the magnetization direction along the static magnetic field.
Finally the Faraday effect creates the optical sidebands, or polarization oscillations of the light induced by the magnetization oscillations \cite{PRB16Hisatomi,PR66Shen,PR67Happer}.
Fig. \ref{fig1}(b) describes the schematic diagram for coherent conversion from microwave photons to optical photons. 
The total Hamiltonian($\hat{H}_t$) describing the conversion process including the KM and a higher MSM can be given by
\begin{equation}
\begin{split}
\hat{H}_{t} &=\hat{H}_{c}+\hat{H}_{s}+\hat{H}_{o},\\
\hat{H}_{c}&=-i\sqrt{2\kappa_{e}}\left[\hat{a}^{\dagger}\hat{a}_{i}(t)-\hat{a}_{i}^{\dagger}(t)\hat{a}\right],\\
\hat{H}_{s}&=\omega_{c}\hat{a}^{\dagger}\hat{a}+\omega_{K}\hat{s}_{K}^{\dagger}\hat{s}_{K} + \omega_{M}\hat{s}_{M}^{\dagger}\hat{s}_{M} +g_{K}\left(\hat{a}^{\dagger}\hat{s}_{K}+\hat{a}\hat{s}_{K}^{\dagger}\right)\\
&\;\;\;+g_{M}\left(\hat{a}^{\dagger}\hat{s}_{M}+\hat{a}\hat{s}_{M}^{\dagger}\right),\\
\hat{H}_{o}&=-i\sqrt{2\delta_{K}}\left(\hat{s}_{K}+\hat{s}_{K}^{\dagger}\right)\left[\hat{b}_{i}(t)e^{i\Omega_{0}t}-\hat{b}_{i}^{\dagger}(t)e^{-i\Omega_{0}t}\right]\\
&\;\;\;-i\sqrt{2\delta_{M}}\left(\hat{s}_{M}+\hat{s}_{M}^{\dagger}\right)\left[\hat{b}_{i}(t)e^{i\Omega_{0}t}-\hat{b}_{i}^{\dagger}(t)e^{-i\Omega_{0}t}\right],
\label{eq1}
\end{split}
\end{equation}
where $\hbar=1$. The subscripts $K$ and $M$ stand for the KM and a higher MSM, respectively.
The MO coversion proceeds with the following steps in the Hamiltonians: $\hat{H}_{c}\rightarrow \hat{H}_{s} \rightarrow \hat{H}_{o}$.
$\hat{H}_{c}$ describes the interaction Hamiltonian between an itinerant microwave mode $\hat{a}_i$ and the microwave cavity mode $\hat{a}$ with external coupling rate $\kappa_{e}$, which results from the rotating-wave approximation.
$\hat{H}_{s}$ including the system Hamiltonian describes the interaction Hamiltonian between cavity and magnon modes.
$g_{K}$ and $g_{M}$ represent the magnon-microwave photon coupling strengths for KM and MSM, respectively. Here, we note that $g_{K}$ and $g_{M}$ include the overlapping coefficient $\xi$, which is related to the space variation effect between the magnetic field of the cavity mode and magnon modes \cite{PRL14Zhang}.
$\hat{a}^{\dagger}\left(\hat{a}\right)$ is the creation (annihilation) operator for the microwave photon at the angular frequency $\omega_{c}$.
$\hat{s}_{K}^{\dagger}\left(\hat{s}_{K}\right)$ and $\hat{s}_{M}^{\dagger}\left(\hat{s}_{M}\right)$ represent the collective spin excitations for KM and MSM at angular frequency $\omega_{K}$ and $\omega_{M}$, respectively (see Appendix A).
The number of spins in a YIG sphere can contribute to both KM and MSM.
$\hat{H}_{o}$ describes the interaction Hamiltonian between magnon modes $\hat{s}$ and a traveling optical photon mode $\hat{b}_i$ with angular frequency $\Omega_{0}$. $\delta_{K}$ and $\delta_{M}$ refer to the optical photon-magnon coupling rate for KM and MSM, respectively.  
Since it includes both Stokes and anti-Stokes processes that are involved in the MO conversion, we leaves the Hamiltonian $\hat{H}_{o}$ without the rotating-wave approximation.
The MO conversion indicates that the itinerant microwave photon mode $\hat{a}_{i}$ is converted to the traveling optical photon mode $\hat{b}_{o}$.

In our experiment, the strongly coupled magnons and cavity microwave photon mode can be determined by normal mode splittings in transmission spectra which are measured from the input and output channels (see Fig. 2(c)). According to the input-output relation \cite{Walls94}, equations of motions for a cavity mode and magnon modes can be obtained from quantum Langevin equation (see Appendix A). As a result, the transmission for multi-magnon modes can be given by
\begin{equation}
S_{21}(\omega)=-i\frac{2\kappa_{e}}{\omega-\omega_{c}+i\kappa_{t}-\sum_{m}g_{m}^{2}\chi_{m}},
\label{eq2}
\end{equation}
where
\begin{equation}
\chi_{m}(\omega)=\left[\omega-\omega_{m}+i\gamma_{m}\right]^{-1}.
\label{eq3}
\end{equation}
Here, $\kappa_{t}=\kappa_{e}+\kappa_{i}$ is the total loss rate which includes both external coupling rate $\kappa_{e}$ and internal loss of the cavity $\kappa_{i}$.

For the conversion process from the microwave to optical wave, the conversion coefficients with amplification factor $\beta_{m}$ for the KM and a MSM are obtained as (see Appendix A)
\begin{equation}
S_{31,K}(\omega)=-2\sqrt{\beta_{K}\delta_{K}\kappa_{e}}\frac{g_{K}\chi_{K}\chi_{c}\left(1+g_{M}^{2}\chi_{M}\chi_{c}T_{M}\right)}{1-g_{K}^{2}\chi_{K}\chi_{c}\left(1+g_{M}^{2}\chi_{M}\chi_{c}T_{M}\right)},
\label{eq4}
\end{equation}
and
\begin{equation}
S_{31,M}(\omega)=-2\sqrt{\beta_{M}\delta_{M}\kappa_{e}}\frac{g_{M}\chi_{M}\chi_{c}\left(1+g_{K}^{2}\chi_{K}\chi_{c}T_{K}\right)}{1-g_{M}^{2}\chi_{M}\chi_{c}\left(1+g_{K}^{2}\chi_{K}\chi_{c}T_{K}\right)},
\label{eq5}
\end{equation}
where
\begin{equation}
\begin{split}
\chi_{c}(\omega)&=\left[\omega-\omega_{c}+i\kappa_{t}\right]^{-1},\\
T_{m}(\omega)&=\left[1-g_{m}^{2}\chi_{m}\chi_{c}\right]^{-1}.
\label{eq6}
\end{split}
\end{equation}
Therefore, the MO conversion efficiency including both KM and a MSM modes can be given by
\begin{equation}
\eta_{t}(\omega)=\left|S_{31,K}(\omega)\right|^{2}+\left|S_{31,M}(\omega)\right|^{2}.
\label{eq7}
\end{equation}
Here, at the resonant condition $\omega-\omega_{c}=\omega-\omega_{K}=\omega-\omega_{M}=0$, the two-mode conversion efficiency can be represented in terms of cooperativities for $\mathcal{C}_{K}=\frac{g_{K}^2}{\kappa_{t}\gamma_{K}}$ and $\mathcal{C}_{M}=\frac{g_{M}^2}{\kappa_{t}\gamma_{M}}$,
\begin{equation}
\eta_{t}=\frac{4\kappa_{e}}{\left(1+\mathcal{C}_{K}+\mathcal{C}_{M}\right)^{2}}\left[\delta_{K}\beta_{K}\frac{\mathcal{C}_{K}^2}{g_{K}^2}+\delta_{M}\beta_{M}\frac{\mathcal{C}_{M}^2}{g_{M}^2}\right].
\label{eq8}
\end{equation}
In this work, the two-mode conversion efficiency given in Eq.(\ref{eq8}) is well matched to the experimental results. If we expand the interaction Hamiltonian to possess higher order terms, the multi-mode MO conversion efficiency can be obtained by
\begin{equation}
\eta_{t}=\frac{4\kappa_{e}}{\left(1+\sum_{m}\mathcal{C}_{m}\right)^{2}}\sum_{m}\left[\delta_{m}\beta_{m}\frac{\mathcal{C}_{m}^2}{g_{m}^2}\right],
\label{eq9}
\end{equation}
where $m$ is a mode index for each MSM. Since so far the MO conversion for multi-modes has not been reported in magnon-cavity system, our theoretical result can be applied to multi-mode conversion based on ferromagnetic material-hybrid systems.

\section{Experiments and Results}
\begin{figure}
\centerline{\scalebox{0.61}{\includegraphics[angle=0]{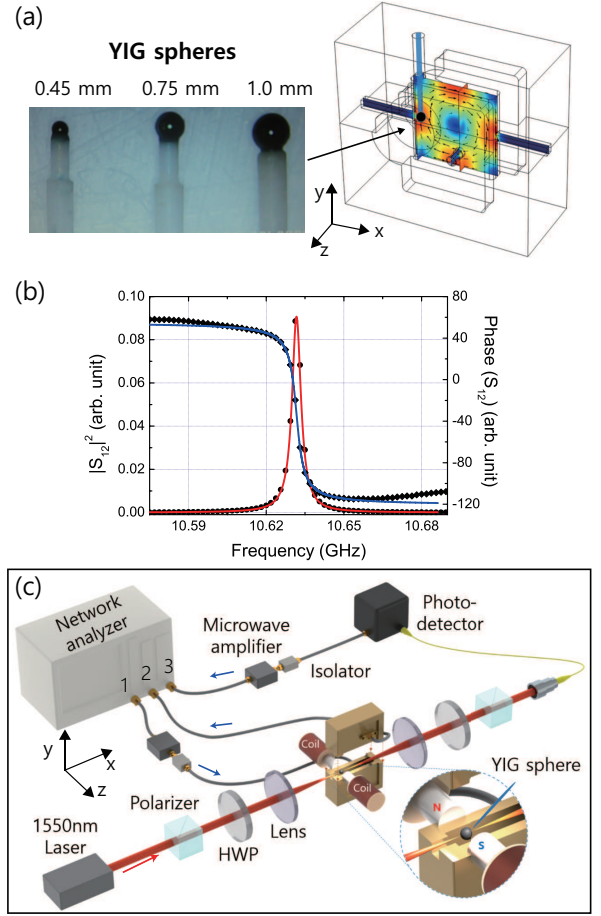}}}
\caption{(a) Numerical simulation of the magnetic field distribution of the fundamental mode TE$_{101}$ inside the microwave cavity with the volume of $20 \times 20 \times 4$ mm$^{-3}$. An YIG sphere made by Ferrisphere Inc. \cite{Ferrisphere} is mounted at the field maximum of the fundamental mode. (b) Transmission power (left y-axis) and phase (right y-axis) without a YIG sphere through the cavity as a function of the microwave frequency. $\omega_{c}/2 \pi=$ 10.632 GHz, $\kappa_{i}/2 \pi=$ 0.6 MHz, $\kappa_{e}/2 \pi=$ 2.1 MHz. Experimental data (black solid-circles and solid-diamonds), and
theoretical results (red and blue curves).
(c) Experimental set-up for the microwave to optical light conversion. To examine the hybrid system, the transmission data are taken by a vector network analyzer. For the conversion measurement, a 1550-nm cw laser with z-polarization is injected into a YIG sphere whose beam waist size is about 120 $\mu$m. The polarization of the light is oscillated by the magnetization oscillations, which is induced by the microwave driving field fed into the channel 1. Finally, the polarization oscillations of light is detected by a fast photodiode and amplified by a microwave amplifier with 30 dB amplification along the channel 3. HWP refers to the half-waveplate.
}
\label{fig2-2}
\end{figure}
\begin{figure*}
\centerline{\scalebox{0.65}{\includegraphics[angle=0]{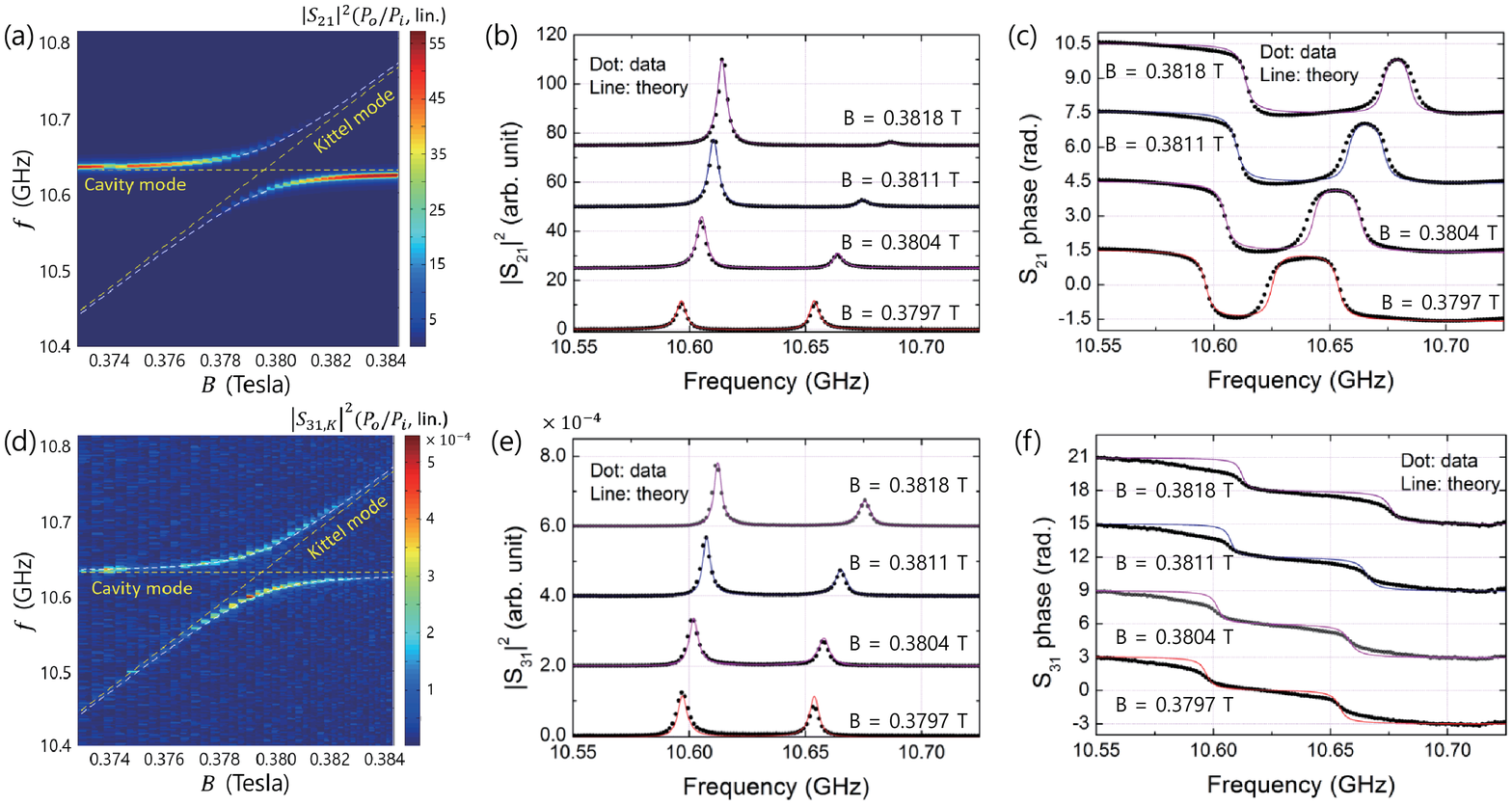}}}
\caption{(a) Measured microwave transmission spectrum, $\left|S_{21} \left( \omega \right) \right|^{2}$ of the 0.45-mm YIG-cavity hybrid system as a function of the microwave frequency and the static magnetic field. The horizontal and diagonal dashed lines (yellow) show the frequency of the fundamental mode TE$_{101}$ and the Kittel mode frequency, respectively. The white-dashed line describes the dispersion of the resonance frequency obtained by diagonalizing $\hat{H}_{s}$ as given in Eq.(\ref{eq1}). (b) Cross sections of $\left|S_{21} \left( \omega \right) \right|^{2}$ at static magnetic fields corresponding to $B=$ 0.3797, 0.3804, 0.3811, and 0.3818 T. Solid lines are theoretical curves given by Eq.(\ref{eq2}) for the data (solid dots). The individual data sets are vertically offset for clarity. (c) The phase $S_{21}(\omega)$ with theoretical hand-fits at the static magnetic fields. (d) Measured MO conversion spectrum, $\left|S_{31,K} \left( \omega \right) \right|^{2}$ of the same system. Here, the MO conversion spectrum is obtained from the raw data which is amplified by a microwave amplifier (30 dB) and detected by a fast photodiode. (e) Cross sections of $\left|S_{31,K} \left( \omega \right) \right|^{2}$ at static magnetic fields corresponding to $B=$ 0.3797, 0.3804, 0.3811, and 0.3818 T. Solid lines are theoretical curves given by Eq.(\ref{eq7}) for the data (solid dots). (f) The phase $S_{31,K}(\omega)$ with theoretical hand-fits at the static magnetic fields. The individual data sets are vertically offset for clarity.
}
\label{fig3}
\end{figure*}

As a ferromagnetic material, we use commercial YIG spheres of diameter 0.45, 0.75, and 1 mm from Ferrisphere and Microsphere. A 3D rectangular cavity is made of oxygen-free copper with the volume $V_{c}$ of $20 \times 20 \times 4$ mm$^{3}$ and the fundamental mode TE$_{101}$  is used for magnetic-dipole coupling. 
Fig. \ref{fig2-2}(a) shows the microwave magnetic field distribution of the fundamental mode TE$_{101}$ at the resonant frequency $\omega_{c}/2 \pi$ of 10.598 GHz, simulated by COMSOL Multiphysics$^{\circledR}$. 
The YIG sphere mounted on the alumina rod along the crystal axis $\left\langle 110 \right\rangle$ is placed near the maximum of the magnetic field in order to get the largest coupling strength and the uniformity of the field as shown in Fig. \ref{fig2-2}(a). 
Fig. \ref{fig2-2}(b) presents measured transmission magnitude and phase without a YIG sphere through the cavity. As a result, the frequency of TE$_{101}$ mode $(\omega_{c}/2 \pi$) is 10.632 GHz, and the external cavity loss rate ($\kappa_{e}/2 \pi$) and the internal cavity loss ($\kappa_{i}/2 \pi$) are 2.1 MHz and 0.6 MHz, respectively.

In order to manipulate the magnetic field, a set of neodymium-iron-boron magnets applies a static magnetic field of around 380 mT to the YIG sphere.
The magnetic components of the microwave field perpendicular to the bias field induce the spin flip, and excite the magnon mode in YIG.
The magnetic circuit consists of a set of permanent magnets and a pair of Helmholtz coils with 800 turns of wires for each.
The  cavity is placed at the center of a pair of Helmholtz coils, so a static magnetic field along the z-axis is applied to the crystal axis $\left\langle 100 \right\rangle$ of the YIG sphere across the cavity. Helmholtz coils driven by a bipolar current supplier combine with the permanent magnets and tunes the resonance frequencies of magnon modes.
The magnetic field measured by a flux gate sensor (3MTS) provides the field-to-current conversion rate of $dB/dI=$ 70 Gauss$/$A. 
Fig. \ref{fig2-2}(c) shows the experimental set-up for the microwave to light conversion. 
We use temperature controlled butterfly diode laser to deliver 1550-nm cw input power of 5 mW before the YIG and get the transmission of 80 \%.
We also use some of optics and microwave components such as polarizer and HWP to define the linear polarization, lens to focus the laser into the YIG, fast photo detector to receive the transmitted laser with optical side band, low noise microwave amplifiers with 30 dB amplification and isolators to increase the signal, and a vector network analyzer by probing the transmission through the hybrid system.

Fig. \ref{fig3}(a) shows the measured microwave transmission spectrum, $\left|S_{21} \left( \omega \right) \right|^{2}$, of the hybrid system with the YIG diameter of 0.45 mm as a function of the frequency and the static magnetic field. A normal-mode splitting is clearly observed, and the avoided crossing manifests strong coupling between the Kittel mode and the microwave cavity mode. As the magnetic field is swept, the Kittel mode approaches the cavity mode up to the degeneracy point. The horizontal dashed line shows the fundamental mode frequency of the cavity and the diagonal dashed line presents the Kittel mode frequency, $f_{11}=\omega_{\mathrm{11}}/2\pi=\mu_{0}\gamma H_{o}/2\pi$ (see Appendix B). White-dashed lines are the dispersion curves of the resonance frequency, $\omega_{\pm}=\frac{\omega_{c}+\omega_{K}}{2}\pm \frac{1}{2}\sqrt{4g_{K}^{2}+(\omega_{c}-\omega_{K})^{2}}$, which are obtained from the diagonalization of the interaction Hamiltonian $\hat{H}_{s}$ in Eq.(\ref{eq1}) without the last term. In order to quantify the coupling strength and the damping rate of ferromagnetic resonace frequency, the experimental data at some of magnetic fields are hand-fitted into the theoretical transmission coefficient $S_{21}\left( \omega \right)$ given in Eq.(\ref{eq2}) (see Fig. \ref{fig3}(b)). As a result, the total cavity linewidth ($\kappa_{t}/2\pi$), the coupling strength ($g_{K}/2\pi$), and the Kittel mode linewidth ($\gamma_{K}/2\pi$) are determined as 2.7, 28.6, and 2.3 MHz, respectively. Here, the coupling strength $g_{K}$ can be represented by $g_{K}=g_{B}\sqrt{2Ns}=\frac{\xi\gamma}{2}\sqrt{\frac{\hbar\omega_{c}\mu_{0}}{V_{c}}}\sqrt{2Ns}$, where $\gamma$ is the electron gyromagnetic ratio, $\omega_{c}$ is the cavity resonance frequency, $V_{c}$ is the volume of the cavity of $20\times20\times4$ mm$^{3}$, and $s=$ 5/2 is the spin number in YIG \cite{PRB16Hisatomi}. $g_{B}$ is the coupling strength of a single Bohr magneton to the cavity which can be calculated as 0.325 Hz for TE$_{101}$ mode in our system. If we assume that all the spins in the YIG sphere are precessing in phase, the coupling strength $g_{K}$ of the Kittel mode to the cavity mode is proportional to the square root of the number of net spins $N_{K}$ \cite{PRL53Agrawal}. In this case, almost all of spins contribute to the KM. The coupling strength $g_{M}/2\pi$ for an MSM is less than 1.0 MHz, so this term can be ignored here. The coefficient $\xi\le1$ indicates the spatial overlap and polarization matching conditions between the microwave and the magnon modes\cite{PRL14Zhang}. From the extracted value of $g_{K}$, we can deduce the number of net spins of $N_{K}=1.51\times10^{17}$. The fact which $g_{K}$ is much larger than $\kappa_{t}$ and $\gamma_{K}$, indicates that the system is in the strong coupling regime even at the room temperature. With the experimental parameters, we obtain a cooperativity of $\mathcal{C}_{K}=g_{K}^{2}/\kappa_{t} \gamma_{K}=$ 132, indicating how well collective spins in the YIG sphere couple to the microwave cavity mode \cite{PRA05Spilane,13Alton}. Fig. \ref{fig3}(c) shows cross-sectional experimental data and theoretical curves for the phases of $S_{21} \left( \omega \right)$. The phase values of $S_{21} \left( \omega \right)$ range from $-\pi/2$ to $\pi/2$ that are two times less than the phase values of $S_{31} \left( \omega \right)$ as shown in Fig. \ref{fig3}(f). In Ref. [26], the phase of a reflection spectrum $S_{11} \left( \omega \right)$ shows the same feature as that of $S_{31} \left( \omega \right)$ except the scale factor of 2. In this work, since all experimental data are based on the $S_{21} \left( \omega \right)$ transmission spectra, the phase of reflection spectrum $S_{11} \left( \omega \right)$ is given in Appendix C. Here, Fig. 7 shows the similar feature to the phase of $S_{31} \left( \omega \right)$. Fig. \ref{fig3}(c) and \ref{fig3}(f) also show that the phase of $S_{21} \left( \omega \right)$ is clearly converted to the phase of $S_{31} \left( \omega \right)$. Therefore, the conveyance of the phase clearly exhibits the coherent conversion from microwave to light.

\begin{figure*}
\centerline{\scalebox{0.61}{\includegraphics[angle=0]{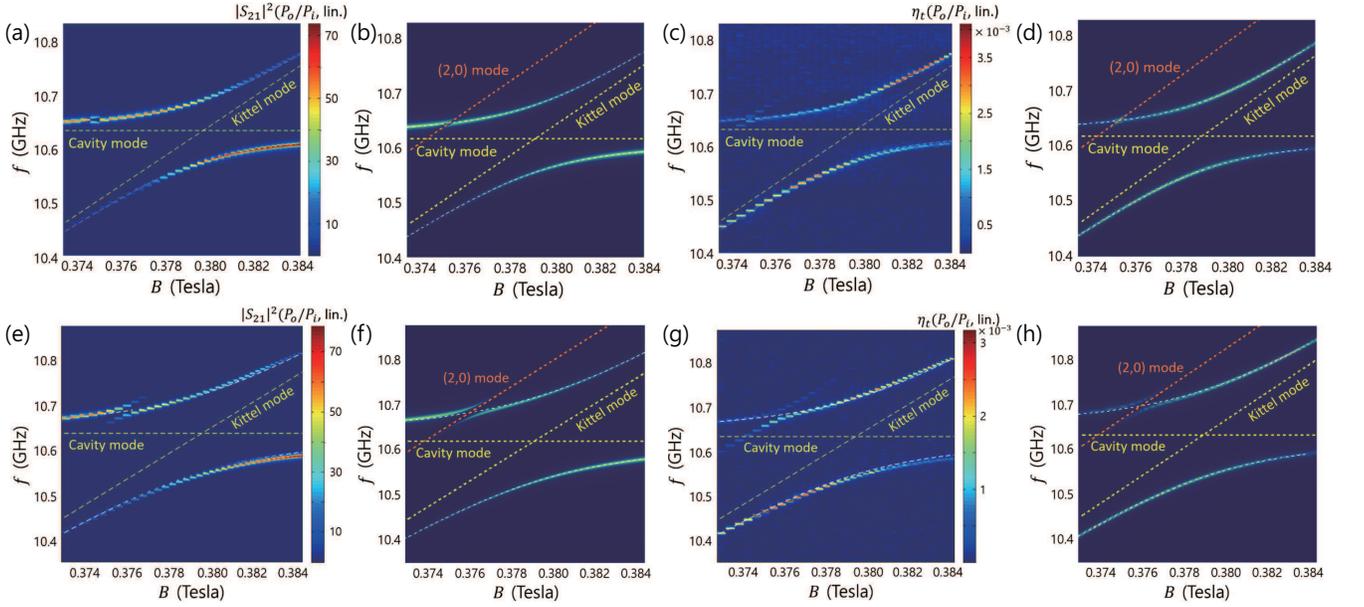}}}
\caption{(a) Measured $\left|S_{21} \left( \omega \right) \right|^{2}$ of the 0.75-mm YIG-cavity hybrid system as a function of the microwave frequency and the static magnetic field. The horizontal and diagonal dashed lines show the frequency of the fundamental mode TE$_{101}$ and the Kittel mode frequency, respectively. The white-dashed line describes the dispersion of the resonance frequency obtained by diagonalizing $\hat{H}_{s}$ as given in Eq.(\ref{eq1}). (b) The simulated spectrum of $\left|S_{21} \left( \omega \right) \right|^{2}$ based on Eq.(\ref{eq2}). For FMR or the Kittel mode, $g_{K}/2\pi$ and $\gamma_{K}/2\pi$ are 67.3 and 1.1 MHz, respectively, and for MSM, $g_{M}/2\pi$ and $\gamma_{M}/2\pi$ are 4 and 1.5 MHz, respectively. The red-dashed line refers to the (2,0) mode. (c) Measured MO conversion spectrum, $\eta_{t}$ of the 0.75-mm YIG-cavity hybrid system. The measured spectrum is based on the raw data which is amplified by a microwave amplifier (30 dB) and detected by a fast photodiode. (d) The simulated spectrum of $\eta_{t}$ of the 0.75-mm YIG-cavity hybrid system based on Eq.(\ref{eq7}). (e) Measured $\left|S_{21} \left( \omega \right) \right|^{2}$ through the 1.0-mm YIG-cavity hybrid system. (f) The simulated spectrum of $\left|S_{21} \left( \omega \right) \right|^{2}$ based on Eq.(\ref{eq2}). For FMR or the Kittel mode, $g_{K}/2\pi$ and $\gamma_{K}/2\pi$ are 91 and 0.95 MHz, respectively, and for MSM, $g_{M}/2\pi$ and $\gamma_{M}/2\pi$ are 12 and 0.9 MHz, respectively. The red-dashed line refers to the (2,0) mode as given in Eq.(\ref{eq11}). (g) Measured $\eta_{t}$ of the 1.0-mm YIG-cavity hybrid system. The measured spectrum is based on the raw data which is amplified by a microwave amplifier (30 dB) and detected by a fast photodiode. (h) The simulated spectrum of $\eta_{t}$ of the 1.0-mm YIG-cavity hybrid system based on Eq.(\ref{eq7}).}
\label{fig4}
\end{figure*}

Fig. \ref{fig3}(d) shows the measured power of the MO conversion coefficient, $\left|S_{31,K}\left( \omega \right)\right|^{2}$, of the hybrid system with the same YIG.
The conversion spectrum $\left|S_{31,K}\left( \omega \right)\right|^{2}$ is almost similar to $\left|S_{21} \left( \omega \right) \right|^{2}$, which implies that the microwave field is coherently converted to the optical wave field. 
Fig. \ref{fig3}(e) shows  the cross sectional experimental data at some of magnetic fields in Fig. \ref{fig3}(d) that are hand-fitted into the theoretical transmission coefficient $\left|S_{31,K}\left( \omega \right)\right|^{2}$ given in Eq.(\ref{eq7}). As a result, $g_{K}/2\pi$ and $\gamma_{K}/2\pi$ are 28.5 MHz and 2.4 MHz, respectively that are quite close to the result of $\left|S_{21} \left( \omega \right) \right|^{2}$. In order to evaluate the MO conversion efficiency, we first estimate the optical photon-magnon coupling rate $\delta_{K}$ which is given by $\delta_{K}=\frac{G_{K}^{2}l^{2}}{16V_{m}}n_{K}\frac{P_{0}}{\hbar\Omega_{0}}$ \cite{PRB16Hisatomi}. With $l=0.45$ mm being the length of the YIG sample, $n_{K}=3.16\times10^{27}$ m$^{-3}$ and $V_{m}=\frac{4}{3}\pi\times0.225^{3}$ mm$^{3}$ being the spin density and the spatial volume of the magnetostatic mode, $\mathcal{V}=3.8$ radians/cm at 1.55 $\mu$m \cite{JMR11Huang} that result in $G_{K}=4\mathcal{V}/{n_{K}}=4.81\times10^{-25}$ m$^{2}$, and $P_{0}/\hbar\Omega_{0}=1.17\times10^{17}$ Hz for $P_{0}=15$ mW, we have $\delta_{K}/2\pi=0.0036$ Hz. Therefore, under the near resonant conditions at $\omega-\omega_{c}=\omega-\omega_{K}=0$, the total conversion efficiency $\eta_{t}=\left|S_{31,K}\right|^{2}$ can be approximated in terms of the coupling strength $g_{K}$,
\begin{equation}
\eta_{t}=\left(\frac{2 \sqrt{\delta_{K}\kappa_{e}}\mathcal{C}_{K}}{g_{K}\left(1+\mathcal{C}_{K}\right)}\right)^{2}.
\label{eq10}
\end{equation}
With all obtained parameters for the YIG sphere with $0.45$ mm diameter, we attain $\eta_{t}=8.45\times10^{-11}$. This low conversion efficiency results from the small light-magnon coupling rate. Actually, the maximum conversion efficiency is occured at particular detunings from the cavity resonance and the Kittel mode frequency \cite{PRB16Hisatomi}. However, in this experiment, we are interested in the multi-mode conversion efficiency at the degenerate point.

In order to examine the YIG size dependence of system parameters, we also measured the transmission spectra of YIG spheres with diameters of 0.75 and 1 mm, as shown in Fig. \ref{fig4}.
Fig. \ref{fig4}(a) and \ref{fig4}(e) show transmission magnitude, $\left|S_{21} \left( \omega \right) \right|^{2}$, measured for YIG diameter of 0.75 and 1.0 mm, respectively. As the size of the YIG sphere increases, the larger number of spins in a bigger sphere can contribute the interaction with the microwave cavity mode, that makes the normal-mode splitting wider. As a result, we obtain the coupling strengths of $g_{K}/2\pi=$ 67.3 and 91 MHz for 0.75 and 1.0-mm YIG spheres so that the cooperativity $C_{K}$ for the Kittel mode reaches up to about 3.5$\times 10^{3}$ as shown in Table \ref{table}. In addition to larger coupling strengths, another avoided level crossing feature is observed because of the strong coupling corresponding to the nonuniform MSMs which can be also coupled to the cavity mode. The coupling strengths of MSM are $g_{M}/2\pi=$ 4 and 12 MHz, and the decay rates are $\gamma_{M}/2\pi=$ 1.1 and 0.9 MHz for YIG spheres with 0.75 and 1.0 mm diameter, respectively. Based on the fitting parameters, 2D spectra of $\left|S_{21} \left( \omega \right) \right|^{2}$ for each case are simulated in Fig. \ref{fig4}(b) and (f). Here, the red-dashed line describes the nonuniform MSM which is identified by magnetostatic theory \cite{PR57Walker,JAP59Fletcher}. In general, the relation between  MSM frequencies and the external magnetic field is linear for $i-\left\vert j \right\vert=0$ or 1 as mentioned in Appendix B. When the YIG sphere is subjected to an oscillating magnetic field at $\omega_{ij}$ and a strong coupling regime is reached at $H_{o}$, avoided level crossings appear at the regions where the resonance frequencies of two subsystems are matched, that make it possible to distinguish a MSM with $i$ and $j$ associated to an avoided level crossing \cite{JAP15Lambert,PRB15Rameshti,JAP16Zhang,PRB20Leo}. In our case, additional avoided level crossing is placed at the (2,0) mode which can be given by \cite{JAP59Fletcher}
\begin{equation}
\omega_{20}=\mu_{0}\gamma M_{s}\sqrt{\left(\frac{H_{o}}{M_{s}}-\frac{1}{3}\right)\left(\frac{H_{o}}{M_{s}}+\frac{7}{15}\right)}.
\label{eq11}
\end{equation}
Fig. \ref{fig4}(c) and (g) show the MO conversion spectra, $\eta_{t}$, measured for the YIG diameter of 0.75 and 1.0 mm, respectively. These MO conversion spectra present raw data which are amplified and detected by using a microwave amplifier and a fast photodiode.
One can find the same avoided level crossing features, as shown in Fig. \ref{fig4}(a) and \ref{fig4}(e), which clearly demonstrates the coherent conversion from microwave to optical photons. Based on the fitting parameters, 2D spectra of $\eta_{t}$ for each case are simulated in Fig. \ref{fig4}(d) and \ref{fig4}(h). As a result, when we take into account both KM and MSM contributions, the total conversion efficiency $\eta_{t}$ are 5.12$\times 10^{-12}$ and 3.46$\times 10^{-12}$ for 0.75 and 1.0 mm YIG spheres, respectively. We summarize system parameters for each YIG sphere in Table \ref{table} that are obtained from the two-mode MO conversion process.

\begin{table}
\caption{System parameters for two-mode MO conversion}
\label{table}
\setlength{\tabcolsep}{4pt}
\begin{tabular}{p{55pt} p{55pt} p{55pt} p{55pt} p{55pt}}
\hline
Parameter&
0.45-mm dia.&
0.75-mm dia.&
1.0-mm dia. \\
\hline
$g_{K}$ [MHz]& 2$\pi\times$ 28.6 & 2$\pi\times$67.3 & 2$\pi\times$91.0 \\
$\gamma_{K}$ [MHz]& 2$\pi\times$2.3 & 2$\pi\times$1.1 & 2$\pi\times$0.95 \\
$g_{M}$ [MHz]& $<$ 2$\pi\times$1.0 & 2$\pi\times$4.0 & 2$\pi\times$12.0 \\
$\gamma_{M}$ [MHz]& $>$ 2$\pi\times$2.0 & 2$\pi\times$1.5 & 2$\pi\times$0.9 \\
$\mathcal{C}_{K}$& 132 & 1373 & 3487 \\
$\mathcal{C}_{M}$& 0.19 & 3.6 & 64 \\
$N_{K}$& 1.51$\times$10$^{17}$ & 8.36$\times$10$^{17}$ & 1.53$\times$10$^{18}$ \\
$N_{M}$& $<$ 1.84$\times$10$^{14}$ & 2.95$\times$10$^{15}$ & 2.66$\times$10$^{16}$ \\
$V_{m}$ [m$^3$]& 4.77$\times$10$^{-11}$ & 2.21$\times$10$^{-10}$ & 5.24$\times$10$^{-10}$ \\
$n_{K}$ [m$^{-3}$]& 3.16$\times$10$^{27}$ & 3.79$\times$10$^{27}$ & 2.92$\times$10$^{27}$ \\
$n_{M}$ [m$^{-3}$]& $<$ 3.87$\times$10$^{24}$ & 1.34$\times$10$^{25}$ & 5.07$\times$10$^{25}$ \\
$G_{K}$ [m$^{2}$] & 4.81$\times$10$^{-25}$ & 4.02$\times$10$^{-25}$ & 5.21$\times$10$^{-25}$ \\
$G_{M}$ [m$^{2}$] & $>$ 3.93$\times$10$^{-22}$ & 1.14$\times$10$^{-22}$ & 2.99$\times$10$^{-23}$ \\
$\delta_{K}$ [mHz] & 2$\pi\times$3.61 & 2$\pi\times$1.80 & 2$\pi\times$1.75 \\
$\delta_{M}$ [Hz] & 2$\pi\times$2.95 & 2$\pi\times$0.512 & 2$\pi\times$0.101 \\
$\eta_{t}$& 8.45$\times$10$^{-11}$ & 5.12$\times$10$^{-12}$ & 3.46$\times$10$^{-12}$ \\
\hline
\multicolumn{4}{p{245pt}}{Subscripts $K$ and $M$ denote the Kittel mode and MSM, respectively.} \\
\end{tabular}
\label{tab1}
\end{table}
\begin{figure}
\centerline{\scalebox{0.8}{\includegraphics[angle=0]{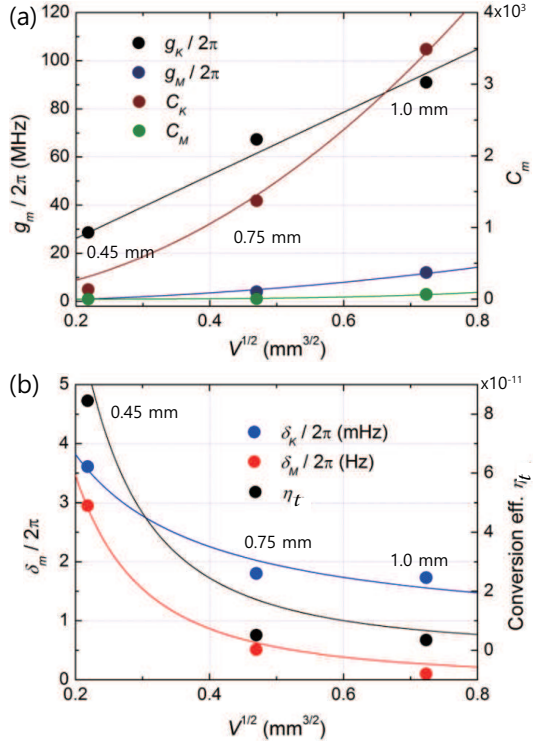}}}
\caption{(a) Extracted values (solid circles) and fit curves for coupling strengths and cooperativities for the KM and MSM as a function of the square root of the YIG volume. (b) Extracted values (solid circles) and fit curves for optical photon-magnon coupling rates for the KM and MSM and the total MO conversion efficiency as a function of the square root of the YIG volume.}
\label{fig5}
\end{figure}

To examine the size dependence of a YIG sphere for MO conversion efficiency, we first evaluate the volume dependence of parameters used for ML conversion efficiency at resonant conditions as shown in Fig. \ref{fig5}. According to the Ref. [30], it demonstrated that the coupling strength $g_{K}$ of the Kittel mode is proportional to the square root of the volume (or the number of spins) of YIG spheres. $g_{K}$ is linear-fitted to $f(x)=130.97x$, where $x$ is the square root of volume $V^{1/2}$. For the higher MSM, $g_{M}$ is not proportional to the linear function, but rather the quardratic function which is  $f(x)=22.19 x^{2}$ (Fig. \ref{fig5}(a)). This might be due to the fact that the spin excitations induced by non-uniform field do not linearly contribute to a higher mode. According to the relation of $\mathcal{C}_{m}=\frac{g_{m}^2}{\kappa_{t}\gamma_{m}}$, the cooperativity $\mathcal{C}_{K}$ is fitted to $f(x)=6569.43 x^{2}$ and for $\mathcal{C}_{M}$, $f(x)=228.79 x^{4}$ is used.

\begin{figure*}
\centerline{\scalebox{0.78}{\includegraphics[angle=0]{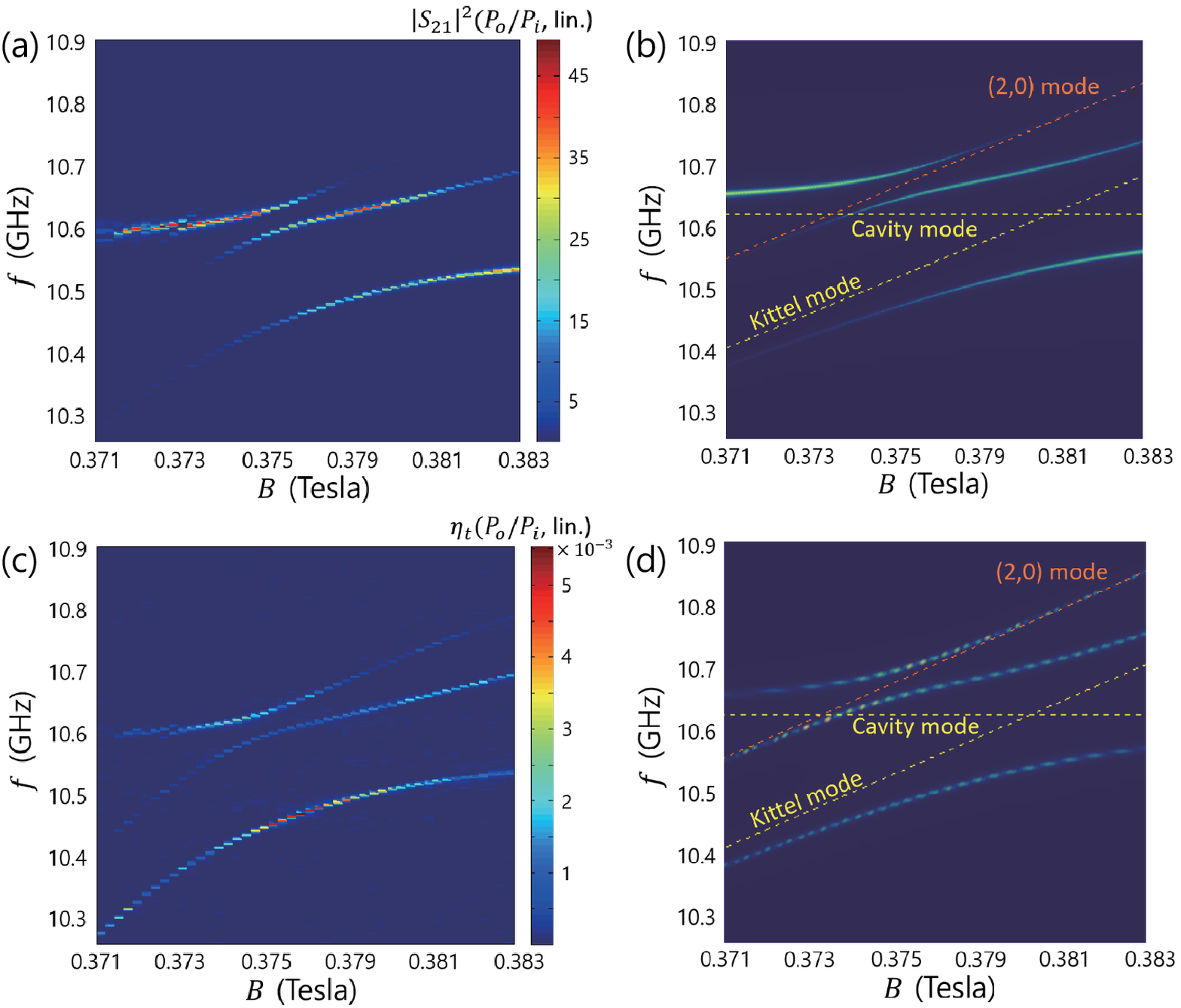}}}
\caption{(a) Measured $\left|S_{21} \left( \omega \right) \right|^{2}$ through the 1.0-mm YIG-cavity hybrid system while YIG position adjusted to enhance the MSMs. (b) The simulated spectrum of $\left|S_{21} \left( \omega \right) \right|^{2}$ from Eq.(\ref{eq2}). (c) Measured $\eta_{t}$ of the same system. The measured spectrum is based on the raw data which is amplified by a microwave amplifier (30 dB) and detected by a fast photodiode. (d)The simulated spectrum of $\eta_{t}$ from Eq.(\ref{eq7}).}
\label{fig6}
\end{figure*}
In addition, $\delta_{K}$ and $\delta_{M}$ also have the dependence of the number of spins. $\delta_{K}$ is fitted to $f(x)=0.693+0.625/x$ and $\delta_{M}$ is fitted to $f(x)=0.139/x^{2}$ as shown in Fig. \ref{fig5}(b). By using these fitting values of system parameters, we obtain the theoretical fit curve for the MO conversion efficiency based on Eq.(\ref{eq8}) as presented in Fig \ref{fig5}(b). As a YIG size increases, the total MO conversion efficiency at the resonant condition decreases since the increments of coupling strength and cooperativity lead to the drop in the MO conversion efficiency as given by Eq.(\ref{eq8}). In our system, the conversion efficiency at the resonant condition is limitted to $10^{-11}$ order. This mainly comes from the small coupling rate $\delta_{K}$ and $\delta_{M}$ between the optical photons and magnons although it depends on the experimental conditions such as the quality of the sample and proper alignment. Therefore, we need to improve the coupling rate $\delta_{m}$ to enhance coherent quantum conversion efficiency between microwave and optical photons. There are several suggestions as mentioned in ref. [26]. One possible way was to use the optical whispering gallery modes (WGMs) of an YIG sphere \cite{PRL16Osada,PRL16Zhang}. No one has achieved a significant improvement, however, supposedly due to the small overlap between the Kittel mode and WGMs. Other suggestions are to utilize a magnetic material with a large Verdet constant such as CrBr$_{3}$ and iron garnet based on rare-earth atoms \cite{Nature11Clausen,Nature11Saglamyurek,JAC04Tsidaeva,PRL19Everts}.

\section{Discussion}
We clearly observe not only the YIG size dependence of the MO conversion but also the coupling strength between the multi-magnetostatic mode and a cavity. 
But note that the multi-mode MO conversion features are not prominent compared to the single-mode MO conversion since the most of spins are involved in the KM mode that makes $g_{M}$ and $\mathcal{C}_{M}$ much lower than the values of $g_{K}$ and $\mathcal{C}_{K}$. 
In order to make the dominant contribution of spins to higher MSM, we carefully position a 1.0 mm-YIG sphere off the uniform microwave mode region, so that a non-uniform MSM also apprears at the degenerate point as shown in Fig. \ref{fig6}(a). In this configuration, the anti-crossings due to the higher MSM become larger since the number of spins contributing to the higher mode increases. Fig. \ref{fig6}(b) shows the simultion result of $\left|S_{21}\left( \omega \right)\right|^{2}$ based on Eq.(\ref{eq2}). As a result, $g_{K}/2\pi$ and $g_{M}/2\pi$ are 83.4 and 25 MHz, respectively, which are few orders larger than decay rates of $\gamma_{K}/2\pi=$1.1 and $\gamma_{M}/2\pi=$0.5 MHz, that indicates the strong couplings between the cavity mode and the KM and MSM. Fig. \ref{fig6}(c) presents the 2D spectrum of $\eta_{t}$. Based on the measured data, $\eta_{t}$ is simulated by using Eq.(\ref{eq7}) as shown in Fig. \ref{fig6}(d). We find out that the theoretical model is well matched with the experimental results. Here, we ignore higher modes in the tail of the spectrum because their contributions are very small in the MO conversion efficiency. The total multi-mode MO conversion efficiency is 1.02 $\times$10$^{-11}$ at the resonant condition. To date, adjustable MO conversion for multi-modes has not been reported in magnon-cavity system.

\section{Conclusion}
We have experimentally demonstrated coherent multi-mode conversion from microwave to optical fields via a YIG sphere in a rectangular microwave cavity. A large number of spins in ferromagnetic materials easily couple the collective excitation to cavity photons, that makes it possible to hybridize the microwave photon modes and magnetostatic modes. A traveling optical field is coupled to a microwave field through this hybrid system. We first observed YIG size dependence of conversion efficiency by measuring the normal-mode splitting between the  magnetostatic modes and the microwave cavity modes, where the coupling strength is in the order of magnitude larger than the decay rates. Based on our multi-mode MO conversion model, we analyzed all the system parameters with experimental data, confirming that the theoretical model is consistent with the experimental results. The total multi-mode conversion efficiency at the resonant condition reaches 1.02$\times 10^{-11}$ for 1.0 mm-YIG sphere. We also evaluate the multi-mode MO conversion efficiency by manipulating position of the YIG sphere inside the cavity.
These sharp and adjustable multi-mode conversion shows the possibility of coherent conversion of multi-mode quantum states while keeping coherence time. This work will also provide optimal design conditions of a cavity magnon-microwave photon system that can be used for coherent conversion between microwave and optical photons.

\begin{acknowledgments}
This work was supported by a grant to Quantum Frequency Conversion Project funded by Defense Acquisition Program Administration and Agency for Defense Development. We would like to thank prof. Changsuk Noh for his valuable comment on the quantum input-output relation and also thank Jinwon Yoo, prof. Wonmin Son, and prof. Suyeon Cho for productive discussion on the hybrid system. 
\end{acknowledgments}

\appendix
\section{The interaction Hamiltonian for the multi-mode microwave-to-optical wave conversion}

The Hamiltonian for the magnon-cavity system can be written as 
\begin{equation}
H_{s}=\omega_{c}\hat{a}^{\dagger}\hat{a}+\sum_{m=K,M}\left[g\mu_{B}B_{z}^{m}\hat{S}_{z}^{m}+g_{m}(\hat{a}\hat{S}_{+}^{m}+\hat{a}^{\dagger}\hat{S}_{-}^{m})\right],
\label{a1}
\end{equation}
where $\omega_{c}$ is the angular frequency of the cavity mode TE$_{101}$, $\hat{a}^{\dagger}$ ($\hat{a}$) is the microwave photon creation (anihilation) operator, and $m=K,M$ denotes the Kittel mode (KM) and magneto static modes (MSM), respectively. $g$ is the electron g-factor, $\mu_{B}$ is the Bohr magneton, and $B_{z}^{m}$ is the effective magnetic field affected by the magnon modes of the YIG sphere. The exchange interaction between electron spins can be ignored because of the long-wavelength descrete modes of spins in the YIG sphere. Since the frequency of the corresponding magnon mode is different from each other, the Hamiltonian for each magnon mode can be written seperately. Here, $\hat{\mathbf{S}}^{m}$ is the collective spin operator for magnon modes which is given by ($\hat{S}_{x}^{m}$,$\hat{S}_{y}^{m}$,$\hat{S}_{z}^{m}$). These collective spin operators can be expressed in terms of the bosonic operators $\hat{s}_{m}^{\dagger}$, $\hat{s}_{m}$ by using the Holstein-Primakoff transformation \cite{PR40Holstein,CRP16Tabuchi}: $\hat{S}_{+}^{m}=\hat{S}_{x}^{m}+i\hat{S}_{y}^{m}=\hat{s}_{m}^{\dagger}\sqrt{2S^{m}-\hat{s}_{m}^{\dagger}\hat{s}_{m}}$, $\hat{S}_{-}^{m}=\hat{S}_{x}^{m}-i\hat{S}_{y}^{m}=(\sqrt{2S^{m}-\hat{s}_{m}^{\dagger}\hat{s}_{m}})\hat{s}_{m}$, and $\hat{S}_{z}^{m}=\hat{s}_{m}^{\dagger}\hat{s}_{m}-2S^{m}$, where $S^{m}$ is the total spin number of the corresponding magnon mode. For the low-lying excitations $\left\langle \hat{s}_{m}^{\dagger}\hat{s}_{m} \right\rangle \ll 2S^{m}$, the Hamiltonian $H_{s}$ can be obtained as
\begin{equation}
H_{s}=\omega_{c}\hat{a}^{\dagger}\hat{a}+\sum_{m=K,M}\left[\omega_{m}\hat{s}_{m}^{\dagger}\hat{s}_{m}+g_{m}(\hat{a}\hat{s}_{m}^{\dagger}+\hat{a}^{\dagger}\hat{s}_{m})\right],
\label{a2}
\end{equation}
where $\omega_{m}=g\mu_{B}B_{z}^{m}$ is the angular frequency of the corresponding magnon mode and $g_{m}=g_{B}\sqrt{2S^{m}}=\frac{\xi\gamma}{2}\sqrt{\frac{\hbar\omega_{c}\mu_{0}}{V_{c}}}\sqrt{2S^{m}}$. Here, $\gamma$ is the electron gyromagnetic ratio, $\omega_{c}$ is the cavity resonance frequency, $V_{c}$ is the volume of the cavity, $g_{B}$ is the coupling strength of a single Bohr magneton to the cavity for TE$_{101}$ mode, and $\xi$ is the spatial overlapping coefficient which is relevant to the spatial variation effect. In the Kittel mode, since the magnetic dipiole coupling between the spins engenders a uniform demagnetization field parallel to the magnetization in a sphere, the demagnetizing field plays no role in the magnetization dynamics for the Kittel mode. For the non-uniform profile for MSM, the variation in space plays a crucial role not only in the frequency calculation but also in the coupling with the exciting field as well as the light.

Therefore, the interaction Hamiltonian of the multi-mode MO conversion can be given by Eq.(\ref{eq1}) which consists of the magnon, microwave photon, optical photon, and their interactions. According to the input-output relation \cite{Walls94}, equations of motions for a cavity mode and magnon modes can be obtained from quantum Langevin equation. For the cavity mode $\hat{a}$,
\begin{equation}
\begin{split}
\dot{\hat{a}}(t)&=-i[\hat{a},\hat{H}_{s}]-\kappa_{t}\hat{a}(t)-\sqrt{2\kappa_{e}}\hat{a}_{i}(t)\\
&=-i\omega_{c}\hat{a}(t)-i\left(g_{K}\hat{s}_{K}(t)+g_{M}\hat{s}_{M}(t)\right)-\kappa_{t}\hat{a}(t)\\
&\;\;\;\;-\sqrt{2\kappa_{e}}\hat{a}_{i}(t),
\label{a3}
\end{split}
\end{equation}
where the total loss rate $\kappa_{t}=\kappa_{e}+\kappa_{i}$ includes both external coupling and internal losses of the cavity. The cavity mode $\hat{a}$ can be given by

\begin{equation}
\hat{a}(t)=\chi_{c}(\omega)\left[g_{K}\hat{s}_{K}(t)+g_{M}\hat{s}_{M}(t)-i\sqrt{2\kappa_{e}}\hat{a}_{i}(t)\right],
\label{a4}
\end{equation}
where
\begin{equation}
\chi_{c}(\omega)=\left[\omega-\omega_{c}+i\kappa_{t}\right]^{-1}.
\label{a5}
\end{equation}
Since magnons do not couple directly to the cavity, no additional input and output magnons are involved. Therefore, the equation of motion of $\hat{s}_{K}$ can be given by
\begin{equation}
\begin{split}
\dot{\hat{s}}_{K}(t)&=-i[\hat{s}_{K},\hat{H}_{s}]-\gamma_{K}\hat{s}_{K}(t)\\
&=-i\omega_{K}\hat{s}_{K}(t)-ig_{K}\hat{a}(t)-\gamma_{K}\hat{s}_{K}(t)\\
\hat{s}_{K}(t)&=\chi_{K}(\omega)g_{K}\hat{a}(t),
\label{a6}
\end{split}
\end{equation}
where
\begin{equation}
\chi_{K}(\omega)=\left[\omega-\omega_{K}+i\gamma_{K}\right]^{-1}.
\label{a7}
\end{equation}
In the same manner, $\hat{s}_{M}$ has the similar form which is
\begin{equation}
\hat{s}_{M}(t)=\chi_{M}(\omega)g_{M}\hat{a}(t),
\label{a8}
\end{equation}
where
\begin{equation}
\chi_{M}(\omega)=\left[\omega-\omega_{M}+i\gamma_{M}\right]^{-1}.
\label{a9}
\end{equation}
Substituting Eq. (\ref{a6}) and Eq. (\ref{a8}) into Eq. (\ref{a4}) and applying the Fourier transform of the cavity mode $\hat{a}$, we can derive
\begin{equation}
\hat{a}(\omega)=-i\sqrt{2\kappa_{e}}T^{-1}\hat{a}_{i}(\omega),
\label{a10}
\end{equation}
where
\begin{equation}
T=\omega-\omega_{c}+i\kappa_{t}-(g_{K}^{2}\chi_{K}+g_{M}^{2}\chi_{M}).
\label{a11}
\end{equation}
In our experiment, we obtain the transmission spectrum which can be determined by measuring the output port 2 from the input port 1. 
For no input in port 2 and the same external coupling rate at both ports, the boundary condition becomes $\hat{a}_{o,2}(\omega)=\sqrt{2\kappa_{e}}\hat{a}(\omega)$ that results in the transmission,
\begin{equation}
S_{21}=\frac{\hat{a}_{o,2}}{\hat{a}_{i,1}}=-i2\kappa_{e}T^{-1}.
\label{a12}
\end{equation}
For the transmission for multi modes, Eq. (\ref{a12}) can be extended to $T=\omega-\omega_{c}+i\kappa_{t}-\sum_{m}g_{m}^{2}\chi_{m}$.

In the conversion process from microwave to optical wave, we can obtain the equation of motions for magnon modes which are given by
\begin{equation}
\begin{split}
\dot{\hat{s}}_{K}(t)=&-i[\hat{s}_{K},\hat{H}_{s}]-\gamma_{K}\hat{s}_{K}(t)\\
&-\sqrt{2\delta_{K}}\left(\hat{b}_{i}(t)e^{i\Omega_{0}t}-\hat{b}_{i}^{\dagger}(t)e^{-i\Omega_{0}t}\right)\\
\dot{\hat{s}}_{M}(t)=&-i[\hat{s}_{M},\hat{H}_{s}]-\gamma_{M}\hat{s}_{M}(t)\\
&-\sqrt{2\delta_{M}}\left(\hat{b}_{i}(t)e^{i\Omega_{0}t}-\hat{b}_{i}^{\dagger}(t)e^{-i\Omega_{0}t}\right).
\label{a13}
\end{split}
\end{equation}
As a result, magnon modes $\hat{s}_{K}$ and $\hat{s}_{M}$ are written as
\begin{equation}
\begin{split}
\hat{s}_{K}(t)=&\chi_{K}\left[g_{K}\hat{a}(t)-i\sqrt{2\delta_{K}} \hat{b}_{i}(t)\right]\\
\hat{s}_{M}(t)=&\chi_{M}\left[g_{M}\hat{a}(t)-i\sqrt{2\delta_{M}} \hat{b}_{i}(t)\right].
\label{a14}
\end{split}
\end{equation}
After substituting Eq. (\ref{a4}) into Eq. (\ref{a14}) and applying the Fourier transform, we can obtain magnon modes for KM and MSM which are given by
\begin{equation}
\begin{split}
\hat{s}_{K}(\omega)=&g_{K}g_{M}\chi_{K}\chi_{c}T_{K}\hat{s}_{M}(\omega)-i\sqrt{2\kappa_{e}}g_{K}\chi_{K}\chi_{c}T_{K} \hat{a}_{i}(\omega)\\
&-i\sqrt{2\delta_{K}}\chi_{K}T_{K} \hat{b}_{i}(\omega)\\
\hat{s}_{M}(\omega)=&g_{K}g_{M}\chi_{M}\chi_{c}T_{M}\hat{s}_{K}(\omega)-i\sqrt{2\kappa_{e}}g_{M}\chi_{M}\chi_{c}T_{M} \hat{a}_{i}(\omega)\\
&-i\sqrt{2\delta_{M}'}\chi_{M}T_{M} \hat{b}_{i}(\omega),
\label{a15}
\end{split}
\end{equation}
where
\begin{equation}
\begin{split}
T_{K}(\omega)&=\left[1-g_{K}^{2}\chi_{K}\chi_{c}\right]^{-1}\\
T_{M}(\omega)&=\left[1-g_{M}^{2}\chi_{M}\chi_{c}\right]^{-1}.
\label{a16}
\end{split}
\end{equation}
If we substitute $\hat{s}_{K}$ ($\hat{s}_{M}$) into $\hat{s}_{M}$ ($\hat{s}_{K}$) in Eq. (\ref{a15}), we can obtain the MO conversion coefficients for KM and MSM. For the KM, by consideringthe Stokes $(\Omega=\Omega_{0}-\omega)$ and anti-Stokes $(\Omega=\Omega_{0}+\omega)$ processes and the boundary conditions $\hat{b}_{o}^{\dagger}(\Omega_{0}-\omega)=\hat{b}_{i}^{\dagger}(\Omega_{0}-\omega)+\sqrt{2\delta_{K}}\hat{c}_{K}(\omega)$ and $\hat{b}_{o}(\Omega_{0}+\omega)=\hat{b}_{i}(\Omega_{0}+\omega)+\sqrt{2\delta_{K}}\hat{c}_{K}(\omega)$ \cite{PRB16Hisatomi}, the conversion coefficient for the KM is obtained as
\begin{equation}
\begin{split}
S_{31,K}(\omega)&=\frac{\sqrt{\beta_{K}}}{2i}\left(\left<\frac{\hat{b}_{o}^{\dagger}(\Omega_{0}-\omega)}{\hat{a}_{i}(\omega)}\right>+\left<\frac{\hat{b}_{o}(\Omega_{0}+\omega)}{\hat{a}_{i}(\omega)}\right>\right)\\
&=-2\sqrt{\beta_{K}\delta_{K}\kappa_{e}}\frac{g_{K}\chi_{K}\chi_{c}\left(1+g_{M}^{2}\chi_{M}\chi_{c}T_{M}\right)}{1-g_{K}^{2}\chi_{K}\chi_{c}\left(1+g_{M}^{2}\chi_{M}\chi_{c}T_{M}\right)},
\label{a17}
\end{split}
\end{equation}
where $\beta_{K}$ is the amplification factor. Here, we point out that, if we take into account the MO conversion of only the KM ($g_{M}=0$), Eq. (\ref{a17}) becomes the same result as the single-mode MO conversion coefficient in Ref. [26]. In the same manner, we can induce the MO conversion coefficient for a MSM by using similar boundary conditions and amplification factor of $\beta_{M}$ that is given by
\begin{equation}
S_{31,M}(\omega)=-2\sqrt{\beta_{M}\delta_{M}\kappa_{e}}\frac{g_{M}\chi_{M}\chi_{c}\left(1+g_{K}^{2}\chi_{K}\chi_{c}T_{K}\right)}{1-g_{M}^{2}\chi_{M}\chi_{c}\left(1+g_{K}^{2}\chi_{K}\chi_{c}T_{K}\right)}.
\label{a18}
\end{equation}
Therefore, the conversion efficiency for the two-mode MO conversion for the KM and a MSM can be obtained as
\begin{equation}
\eta_{t}(\omega)=\left|S_{31,K}(\omega)\right|^{2}+\left|S_{31,M}(\omega)\right|^{2}.
\label{a19}
\end{equation}

\section{Magnetostatic modes in a ferromagnetic sphere}
Magnons are spin excitations describing small perturbations to the magnetization of a ferromagnetic system. A small oscillating magnetic field in the plane perpendicular to the bias field can lead the alignment of spins to deviate slightly from the bias direction. 
The bias field exerts a torque on misaligned spins, and then the spins begin precessing around it. L.R. Walker first considered the relationship between the resonance frequency and the internal static field of a ferromagnetic spheroid \cite{PR57Walker,PR58Walker}. 
He assumed that the microwave magnetic fields in spheroids satisfy the magnetostatic approximations. 
The allowed resonant frequencies of MSMs in a sphere inserted in a microwave cavity can be derived from the characteristic equation in terms of associated Legendre function $P_{i}^{j}(\xi_{0})$ \cite{JAP59Fletcher}.
\begin{equation}
i+1+\xi_{0}\frac{{P_{i}^{j}}^{\prime}(\xi_{0})}{P_{i}^{j}(\xi_{0})}\pm j\chi_{2}=0,
\label{b1}
\end{equation}
where $\xi_{0}^{2}=1+\frac{1}{\chi_{1}}$, $\chi_{1}=\frac{\gamma^{2}M_{s}H_{i}}{\gamma^{2}H_{i}^{2}-f^{2}}$, $\chi_{2}=\frac{\gamma M_{s}f}{\gamma^{2}H_{i}^{2}-f^{2}}$, $H_{i}=H_{0}-\frac{M_{s}}{3}$, and ${P_{i}^{j}}^{\prime}(\xi_{0})=\frac{dP_{i}^{j}(\xi_{0})}{d\xi_{0}}$. Here, $H_{i}$ and $H_{o}$ are internal and external magnetic fields, respectively. $\mu_{0}M_{s} = 0.178$ T (at 298 K) \cite{09Stancil,JAP74Hansen} is the saturation magnetization, $\mu_{0}$ is the vacuum permeability, $\frac{\gamma}{2\pi} = 28$ GHz/T is the gyromagnetic ratio, and $f$ is the frequency. $i$ and $j$ are mode indices that $i\ge1$ is an integer and $j$ is also an integer obeying $-i\le j \le i$.

\begin{figure*}
\centerline{\scalebox{0.8}{\includegraphics[angle=0]{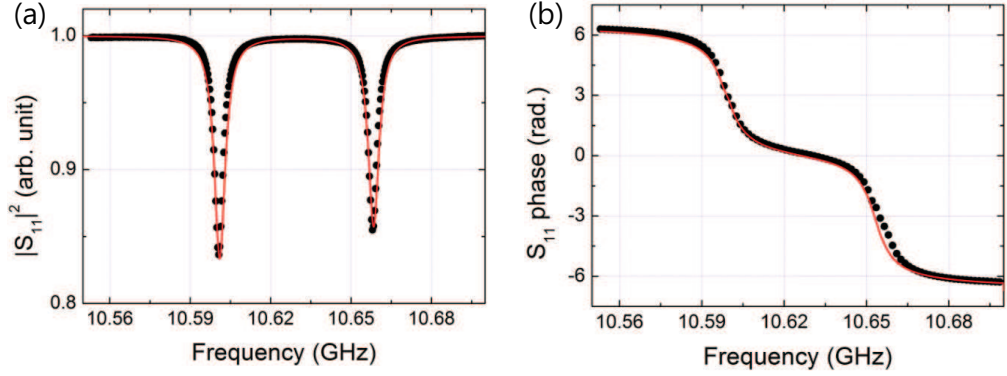}}}
\caption{(a) Reflection coefficient $\left|S_{11}(\omega)\right|^{2}$ of the 0.45 mm-dia YIG-cavity hybrid system as a function of the microwave frequency. (b) The phase $S_{11}(\omega)$ of the 0.45 mm-dia YIG-cavity hybrid system as a function of the microwave frequency. Solid lines are theoretical curves given by Eq.(\ref{c1}).}
\label{fig7}
\end{figure*}

For a single mode solution, it is labelled with $(i,j)$. For MSMs with $i-\left\vert j \right\vert=0$ or 1, the relations between the resonant frequencies and the external magnetic field can be given by \cite{JAP59Fletcher,PSSB77Roschmann}
\begin{equation}
\frac{\omega_{ij}}{\mu_{0}}=\gamma H_{o}+\left(\frac{j}{2j+1}-\frac{1}{3}\right)\gamma M_{s} \quad (i=j),
\label{b2}
\end{equation}
\begin{equation}
\frac{\omega_{ij}}{\mu_{0}}=\gamma H_{o}+\left(\frac{j}{2j+3}-\frac{1}{3}\right)\gamma M_{s} \quad (i=j+1).
\label{b3}
\end{equation}
Here, the $(1,1)$ FMR mode, known as the Kittel mode, is the lowest mode in which all spins precess in phase which gives a frequency given by $\omega_{\mathrm{11}}=\mu_{0}\gamma H_{o}$.

\section{Microwave reflection spectrum}

Figure 7(a) and (b) show the measured reflection spectrum $\left|S_{11}(\omega)\right|^{2}$ and the phase $S_{11}(\omega)$ of the hybrid system with the 0.45 mm-dia YIG as a function of the microwave frequency. From the boundary conditon $\hat{a}_{o}(\omega)=\hat{a}_{i}(\omega)+\sqrt{2\kappa_{e}}\hat{a}(\omega)$ and equations (\ref{a1}) and (\ref{a4}), we can easily obtain the reflection coefficient $S_{11}\left( \omega \right)$,

\begin{equation}
S_{11}(\omega)=1-\frac{i2\kappa_{e}}{\omega-\omega_{c}+i\kappa_{t}-\sum_{m}g_{m}^{2}\chi_{m}}.
\label{c1}
\end{equation}

\end{document}